%%%%%%%%%%%%%%%%%%%%%%%%%%%%%%%%%%%%%%%%%%%%%%%%%%%%%
%
%
%      Random hermitian matrices in an external field
%
%                          by
%
%              Paul Zinn-Justin
%
%
%%%%%%%%%%%%%%%%%%%%%%%%%%%%%%%%%%%%%%%%%%%%%%%%%%%%%
%
%
%         TEX-file using PHYZZX-macro
%
%
%
%
%%%%%%%%%%%%%%%%%%%%%%%%%%%%%%%%%%%%%%%%%%%%%%%%%%%%%
%
%
%
\input phyzzx
%
%%%%%%%%%%%%%%%%%%%%%%%%%%%%%%%%%%%%%%%%%%%%%%%%%%%%%%
% This will make your PHYZZX pagesize wider and longer
% It is OPTIONAL. It redefines the papers macro
%
\catcode`\@=11 % This allows us to modify PLAIN macros.
\def\papersize{\hsize=40pc \vsize=53pc \hoffset=0pc \voffset=1pc
   \advance\hoffset by\HOFFSET \advance\voffset by\VOFFSET
   \pagebottomfiller=0pc
   \skip\footins=\bigskipamount \normalspace }
\catcode`\@=12 % at signs are no longer letters

\def\sin{\mathop{\rm sin}\nolimits}

\papers

%%%%%%%%%%%%%%%%%%%%%%%%%%%%%%%%%%%%%%%%%%%%%%%%%%%%%%

\vsize=23.cm
\hsize=15.cm

\tolerance=500000
\overfullrule=0pt

\Pubnum={LPTENS-97/08 \cr
{\tt cond-mat/9703033} \cr
March 1997}

\date={}
\pubtype={}
\titlepage
\title{\bf  Random hermitian matrices in an external field}
\author{P.~Zinn-Justin}
\vskip 1.cm
\address{
Laboratoire de Physique Th\'eorique de l'Ecole
Normale Sup\'erieure
\foot{{\rm unit\'e propre du CNRS, associ\'ee \`a l'Ecole Normale
Sup\'erieure et l'Universit\'e Paris-Sud.}}
 \break 24 rue Lhomond, 75231
Paris Cedex 05, France\break
{\tt pzinn@physique.ens.fr}
} 

\vskip 1.cm
\abstract{
In this article, a model of random hermitian matrices is considered, in which
the measure $\exp(-S)$ contains a general $U(N)$-invariant potential
and an external source term: $S=N\tr(V(M)+MA)$.
The generalization of known determinant formulae
leads to compact expressions for the correlation functions of the
energy levels. These expressions, exact at finite $N$,
are potentially useful for asymptotic analysis.


}

\endpage
\pagenumber=1

\REF\WI{E.~P.~Wigner, {\it Proc. Cambridge Philos. Soc.} 47 (1951) 790,
and other papers reprinted in C.~E.~Porter, {\it Statistical theories of spectra: fluctuations} (Academic Press,
New York, 1991).}
\REF\ME{M.~L.~Mehta, {\it Random matrices}, 2nd ed. (Academic Press,
New York 1991).}
\REF\DY{F.~J.~Dyson, {\it J. Math. Phys.} 13 90 (1972).}
\REF\IZ{Harish~Chandra, {\it Amer. J. Math.} 79 (1957) 87-120. \hfill\break
C.~Itzykson and J.-B.~Zuber, {\it J. Math. Phys.} 21 (1980) 411.}
\REF\KAZ{V.~A.~Kazakov, {\it Nucl. Phys.} B (Proc. Suppl.) 4 (1988) 93.}
\REF\AMB{J.~Ambjorn and Yu.~M.~Makeenko, {\it Mod. Phys. Lett.}
A 5 (1990), 1753.}
\REF\GRN{D.~J.~Gross and M.~J.~Newman, {\it Phys. Lett.} B 266 (1991), 291.}
\REF\BZ{E.~Br\'ezin and A.~Zee, {\it Nucl. Phys.} B 402 (1993), 613.}
\REF\MEL{P.~A.~Mello, {\it Theory of random matrices:
spectral statistics and scattering problems}
in {\it Mesoscopic quantum physics}, Les Houches Session
LXI, E.~Akkermans, G.~Montambaux, J.-L.~Pichard, and J.~Zinn-Justin eds.
(North-Holland, 1994).}
\REF\BEEN{C.~W.~J.~Beenakker, {\it Nucl. Phys.} B 422 (1994), 515.}
\REF\EY{B.~Eynard, {\it Gravitation quantique bidimensionnelle et matrices
al\'eatoires}, th\`ese de doctorat de l'Universit\'e Paris 6 (1995).}
\REF\BH{E.~Br\'ezin and S.~Hikami, {\it Nucl. Phys.} B 479 (1996) 697.\hfill\break
E.~Br\'ezin and S.~Hikami, preprint LPTENS-96/51, UT-Komaba-96/14,
cond-mat/9608116.\hfill\break
E.~ Br\'ezin and S.~Hikami, preprint cond-mat/9702213.}

\chapter{Introduction.}
As first suggested by Wigner [\WI],
random matrices can be used to simulate Hamiltonians of
complex systems.
In this approach, one would like to
characterize the structure of the energy levels, which are represented by the
eigenvalues of a large matrix
that can be assumed random.
It is now known that many statistical properties of spectra
of true physical systems are indeed well described by those of random matrices
(cf [\MEL] for a review): it is therefore important to understand how much these
spectral properties depend on the particular matrix ensemble chosen, i.e. determine
universality classes of matrix ensembles.

We shall consider here ensembles of hermitian matrices only, which
correspond to systems without time-reversal invariance.
Let us introduce the probability distributions $\rho_n$ 
of the eigenvalues: If $M$ is a random hermitian $N\times N$
matrix, we define $\rho_n(\lambda_0,\ldots,\lambda_{n-1})$ to be
the density of probability that $M$ has $(\lambda_0,\ldots,\lambda_{n-1})$
among its $N$ eigenvalues, with the normalization convention that:
$\int\prod_{i=0}^{n-1} d\lambda_i\, \rho_n(\lambda_0,\ldots,\lambda_{n-1})=1.$
Following [\ME], we also define $R_n={N!\over (N-n)!}\rho_n$, 
a different normalization which
makes the connection with the correlation functions
$$\left<\prod_{i=0}^{n-1}\tr\delta(M-\lambda_i)\right>
=R_n(\lambda_0,\lambda_1,\ldots,\lambda_{n-1})\quad\hbox{\rm for distinct }\lambda_i\eqn\cor$$
since these quantities only differ by $\delta$ functions for coinciding eigenvalues.

In the case of a $U(N)$-invariant measure, consisting of a simple potential
term, which we briefly review in section 2, the theory of orthogonal
polynomials allows to write down exact expressions at finite $N$ 
for these functions, in terms of a single kernel $K(\lambda,\mu)$ [\ME]:
$$R_n(\lambda_0,\lambda_1,\ldots,\lambda_{n-1})=
\det (K(\lambda_i,\lambda_j))
_{i,j=0\ldots n-1}.\eqn\detform$$
The study of the distribution of eigenvalues then boils down
to the analysis of this kernel; in particular the asymptotics of $K$
as $N\rightarrow\infty$ allow to compute asymptotics of correlation
functions and find the different ``universal'' properties that arise in this
limit. Therefore it seems quite interesting
to find similar formulae for more general
measures. In fact, in recent papers [\BH],
Br\'ezin and Hikami have shown that formula $\detform$
can be generalized to the
gaussian ensemble with an external field. More precisely,
for the (unnormalized) measure 
$$\exp \left(-{N\over 2} \tr M^2 + N\tr MA\right) d^{N^2}M, \eqn\gaus$$
they introduced the kernel
$$\tilde{K}(\lambda,\mu)={1\over N} \int {dt\over 2\pi} \oint {dv\over 2\pi i}
\prod_{l=0}^{N-1}\left({it/N-a_l\over v-a_l}\right) {1\over it/N-v}
e^{-{N\over 2} v^2 - t^2/2N -it\lambda +Nv\mu}\eqn\bhker$$
(the $\sim$ is here to distinguish this kernel from a slightly different
one that will be introduced later)
where the $a_l$ are the eigenvalues of  the hermitian matrix $A$ and the
contour integral encircles these eigenvalues. $\detform$ then holds
with $\tilde{K}$ instead of $K$.

The aim of this paper is to define a kernel $K$ such that $\detform$
still holds in the more general case of an arbitrary potential
with an external source term,
which is the subject of section 3. As an example we 
consider in section 4 the gaussian ensemble with an external field and
reproduce the kernel $\tilde{K}$ obtained in [\BH].

\chapter{The $U(N)$-invariant case.} Let us consider an ensemble
of random hermitian $N\times N$ matrices with the measure
$$Z^{-1} \exp \left(-N\tr V(M)\right) d^{N^2}M \eqn\meas$$
where $V$ is a polynomial and $Z$ the partition function.
A classical result [\ME] expresses
the distribution law $\rho_n$ of $n$ eigenvalues ($1\le n \le N$) of $M$
in terms of the kernel
$$K(\lambda,\mu)=\sum_{k=0}^{N-1} F_k(\lambda) F_k(\mu).\eqn\stdker$$
Here $F_i$ is the orthonormal function associated to the usual orthogonal
polynomial $P_i(\lambda)=\lambda^i+\cdots$:
$$\eqalign{
&F_i(\lambda)=h_i^{-1/2} P_i(\lambda) \exp\left(-{N\over 2} V(\lambda)\right)\cr
&\int d\lambda \exp(-N V(\lambda)) P_i(\lambda) P_j(\lambda) = h_i \delta_{ij}.\cr
}\eqn\orth$$
(see [\EY] for a review of orthogonal polynomials in matrix models).
Let us briefly rederive this result in a manner that naturally generalizes.
As the measure $\meas$ only depends on the eigenvalues of $M$, the integration
over the angular variables is trivial and one finds:
$$\rho_N(\lambda_0,\lambda_1,\ldots,\lambda_{N-1})=Z^{-1} \Delta^2(\lambda_i)
\exp \left(-N \sum_{i=0}^{N-1} V(\lambda_i)\right).\eqn\rhoN$$
The Van der Monde determinant $\Delta(\lambda_i)=\det (\lambda_i{}^j)
_{i,j=0\ldots N-1}$ can be rewritten in terms of the orthogonal polynomials:
$$\rho_N(\lambda_0,\lambda_1,\ldots,\lambda_{N-1})=Z^{-1} \det (P_k(\lambda_i))
_{i,k=0\ldots N-1} \det (P_k(\lambda_j))_{j,k=0\ldots N-1}
\exp \left(-N \sum_{i=0}^{N-1} V(\lambda_i)\right).\eqn\rhoNb$$
One can now easily compute $Z=N! \prod_{i=0}^{N-1} h_i$ by integrating over
all $\lambda_i$. Combining the two determinants, we finally obtain:
$$\rho_N(\lambda_0,\lambda_1,\ldots,\lambda_{N-1})= {1\over N!}
\det (K(\lambda_i,\lambda_j))
_{i,j=0\ldots N-1}.\eqn\rhoNc$$
The kernel $K$ has the following properties:
$$\left\{\eqalign{
K(\lambda,\mu)&=K(\mu,\lambda)\cr
[K\star K](\lambda,\rho) &\equiv \int d\mu \, K(\lambda,\mu) K(\mu,\rho)
= K(\lambda,\rho) \cr
}\right.\eqn\kerprop$$
i.e. it is the orthogonal projector on the subspace spanned by the $F_k$, $0\le k
\le N-1$. Using the property $K\star K=K$ and noting that
$$\rho_n(\lambda_0,\lambda_1,\ldots,\lambda_{n-1})=\int d\lambda_n
\rho_{n+1}(\lambda_0,\lambda_1,\ldots,\lambda_n),\eqn\induc$$
one can then show inductively that
$$\rho_n(\lambda_0,\lambda_1,\ldots,\lambda_{n-1})={(N-n)!\over N!}
\det (K(\lambda_i,\lambda_j))_{i,j=0\ldots n-1}.\eqn\detformb$$
for any $n\le N$. This is equivalent to formula $\detform$.

\chapter{Generalization to the case of an external field.} 
We shall now see that
in the case of a general measure with an external field, $\detform$ still holds;
a simple expression for a kernel $K$ will be derived. Let us indeed consider the
measure:
$$Z^{-1} \exp \left(-N \tr V(M)+N \tr MA\right) d^{N^2}M \eqn\measb$$
where $V$ is an arbitrary polynomial, and $A={\rm diag}
(a_0,\ldots,a_{N-1})$ can be assumed diagonal.

Particular matrix models of this type and their large $N$ study appear in many papers [\KAZ,\GRN].
Here we shall go beyond the $1/N$-expansion and write exact expressions at finite $N$.

One diagonalizes $M$: if $M=\Omega\Lambda\Omega^\dagger$
where $\Lambda={\rm diag}(\lambda_0,\ldots,\lambda_{N-1})$, 
the integral over $\Omega$ is the usual Itzykson--Zuber integral [\IZ]
on the unitary group and we find:
$$\rho_N(\lambda_0,\lambda_1,\ldots,\lambda_{N-1})= Z^{-1} \Delta(\lambda_i)
{\det (\exp N \lambda_j a_l) \over \Delta(a_l)}
\exp \left(-N \sum_{i=0}^{N-1} V(\lambda_i)\right).\eqn\rhoNd$$
We replace as usual powers of  $\lambda$ in the Van der Monde
with the orthogonal polynomials $P_k(\lambda)$ of the measure 
$\exp -N V(\lambda) d\lambda$.
$Z$ can now be computed:
$$\eqalign{
Z&=N!{1\over \Delta(a_l)} \int \prod_{i=0}^{N-1} d\lambda_i \det(P_k(\lambda_i))
\exp N \sum_{i=0}^{N-1} (-V(\lambda_i) + a_i \lambda_i)\cr
&= {N!\over \Delta(a_l)}\det\left(\int d\lambda P_k(\lambda) \exp N (-V(\lambda)+a_l \lambda)\right)
_{k,l=0\ldots N-1}\cr
} \eqn\Z$$
Inserting $\Z$ into $\rhoNd$ yields
$$\rho_N(\lambda_0,\lambda_1,\ldots,\lambda_{N-1})={1\over N!}
{\det(P_k(\lambda_i))_{i,k=0\ldots N-1}\, \det(\exp N a_l \lambda_j)_{j,l=0\ldots N-1}
\over \det\left(\int d\lambda P_k(\lambda) \exp N (-V(\lambda)+a_l \lambda)\right)
_{k,l=0\ldots N-1}}
\exp \left(-N \sum_{i=0}^{N-1} V(\lambda_i)\right).\eqn\longdet$$
This formula has a remarkably simple structure, which can be made more explicit
by introducing $F_k(\lambda)=h_k^{-1/2}
P_k(\lambda) \exp -{N\over 2} V(\lambda)$ as before, and $G_l(\lambda)=
\exp \left[Na_l \lambda-{N\over 2} V(\lambda)\right]$:
$$\rho_N(\lambda_0,\lambda_1,\ldots,\lambda_{N-1})={1\over N!}
{\det(F_k(\lambda_i))_{i,k=0\ldots N-1}\, \det(G_l(\lambda_j)_{j,l=0\ldots N-1}
\over \det(\int d\lambda F_k(\lambda) G_l(\lambda))_{k,l=0\ldots N-1}}
.\eqn\shortdet$$
The matrix $(\int d\lambda G_l(\lambda) F_k(\lambda))_{l,k=0\ldots N-1}$
possesses an inverse, which we denote by $\alpha_{kl}$; putting
together the three determinants we finally obtain:
$$\rho_N(\lambda_0,\lambda_1,\ldots,\lambda_{N-1})={1\over N!}
\det(K(\lambda_i,\lambda_j))_{i,j=0\ldots N-1}\eqn\detN$$
where
$$K(\lambda,\mu)=\sum_{k,l=0}^{N-1} F_k(\lambda) \alpha_{kl} G_l(\mu).
\eqn\defK$$
The kernel $K$ satisfies the property:
$$\eqalign{
[K\star K](\lambda,\rho)&=\sum_{k,k',l,l'=0}^{N-1} \alpha_{kl} F_k(\lambda)
\left[ \int d\mu G_l(\mu) F_{k'}(\mu) \right]
\alpha_{k'l'} G_{l'}(\rho)\cr
&=K(\lambda,\rho).\cr
%&=\sum_{k,l,l'=0}^{N-1} \alpha_{kl}  F_k(\lambda) \delta_{ll'}G_{l'}(\rho)\cr
}\eqn\kerpropb$$
Thus, one can follow the same line of reasoning as
in the $U(N)$-invariant case to obtain the determinant formulae
$$\rho_n(\lambda_0,\lambda_1,\ldots,\lambda_{n-1})={(N-n)!\over N!}
\det(K(\lambda_i,\lambda_j))_{i,j=0\ldots n-1}\eqn\detformc$$
for any $n\le N$.

The kernel $K$ is of the form
$$K(\lambda,\mu)=\sum_{k=0}^{N-1} F_k(\lambda) \hat F_k(\mu)\eqn\kerform$$
with $\hat F_k(\mu)=\sum_l \alpha_{kl} G_l(\mu)$; but $\hat F_k\ne F_k$ and
therefore $K$ is not symmetric. 
Further analysis of $K$ resides in the observation that as $a\rightarrow 0$,
$$\int d\lambda P_k(\lambda) \exp N (-V(\lambda)+a\lambda)
=h_k {N^k\over k!} a^k + O\left(a^{k+1}\right)\eqn\lapl$$
due to the orthogonality of the $P_k$. The first consequence is that it is only in
the limit $A\rightarrow 0$ that the $G_l$ become linear combinations of the
$F_k$, $0\le k \le N-1$, ensuring that $\hat F_k\rightarrow F_k$, as expected.
The second consequence is that the quantity
$$a^{-N}\prod_{l=0}^{N-1} (a-a_l) 
\int d\lambda P_k(\lambda) \exp N (-V(\lambda)+a\lambda)\eqn\laplb$$
is well-defined for any $k\ge N$; we can define the inverse Laplace transform
$$\Phi_k=\exp {N\over 2} V(\lambda)\, \partial^{-N}
\prod_{l=0}^{N-1} (\partial-a_l)
\left[P_k(\lambda)\exp -N V(\lambda)\right],\eqn\laplc$$
(where $\partial\equiv 1/N\, d/d\lambda$) which is a regular function with fast decrease
at infinity. One can see that we now have a complete set of eigenvectors of $K$:
$$\left\{\eqalign{
\int K(\lambda,\mu) F_k(\mu) d\mu&=F_k(\lambda)\quad\forall k<N\cr
\int K(\lambda,\mu) \Phi_k(\mu) d\mu &= 0 \quad\forall k\ge N.\cr
}\right.\eqn\eigen$$
Again, it is only when $A=0$ that $\Phi_k=F_k$.

We conclude that $K$ is a non-orthogonal projector on the space spanned
by the $F_k$, $0\le k\le N-1$.
 
\chapter{Example: the gaussian ensemble.}
In the case where $V(M)={1\over 2} M^2$, the kernel
$K$ defined in eq.~$\defK$ should be related to the kernel $\tilde{K}$
of [\BH].
This is indeed what we find. 

The orthogonal polynomials $P_k(\lambda)$ are Hermite polynomials, their
normalization is $h_k=\sqrt{2\pi/N} k!/N^k$, and one can compute explicitly the integral
$$\int d\lambda P_k(\lambda) e^{N (-{1\over 2}\lambda^2+a \lambda)}=
\sqrt{2\pi/N} a^k e^{{N\over 2}a^2}\eqn\expl$$
in accordance with eq.~$\lapl$.
One can also directly calculate $Z$ by
changing variables from $M$ to $M-A$ in the gaussian matrix integral:
this immediately gives
$$Z \sim \exp {N\over 2}\sum_{l=0}^{N-1} a_l^2.\eqn\Zgauss$$
up to $A$-independent factors. We deduce the explicit form of the kernel $K$:
$$K(\lambda,\mu)=\sqrt{N\over 2\pi} \sum_{k,l=0}^{N-1} P_k(\lambda) \beta_{kl}
e^{N(a_l\mu-{1\over 4}(\lambda^2+\mu^2)-{1\over 2}a_l^2)}.\eqn\Kgauss$$
where $\beta_{kl}=\alpha_{kl} \sqrt{2\pi/N}\exp {N\over 2}a_l^2$
is the inverse of the Van der Monde matrix $(a_l{}^k)$.
By definition one has:
$$\sum_{k=0}^{N-1} \beta_{kl} a^k
= \prod_{l'\ne l} {a-a_{l'}\over a_l-a_{l'}}.\eqn\vander$$
which allows to express $\beta_{kl}$ in terms of symmetric functions of the $(a_{l'})_{l'\ne l}$.
We now use an inverse formula of eq.~$\expl$:
$$P_k(\lambda)=\int_{-i \infty}^{+i\infty} {da\over i \sqrt{2\pi N}} a^k
e^{N({1\over 2} a^2-a\lambda + {1\over 2} \lambda^2)}\eqn\invexpl$$
where the integration is along the imaginary axis.
Combining eq.~$\Kgauss$, $\vander$, and $\invexpl$ we obtain:
$$K(\lambda,\mu)=\sum_{l=0}^{N-1} \int_{-i\infty}^{+i\infty} {da\over 2\pi i}
\prod_{l'\ne l}{a-a_{l'} \over a_l-a_{l'}}
e^{N({1\over 2} a^2-a\lambda+{1\over 4}\lambda^2-{1\over 4}\mu^2
-{1\over 2}a_l^2 + a_l\mu)}.
\eqn\kergauss$$
This can in turn be represented by a contour integral in the complex plane which
picks up poles at $v=a_l$:
$$K(\lambda,\mu)=\int_{-i\infty}^{+i\infty} {da\over 2\pi i}
\oint {dv\over 2\pi i} \prod_{l=0}^{N-1}\left({a-a_l \over v-a_l}\right)
{1\over a-v}
e^{N({1\over 2} a^2-a\lambda+{1\over 4}\lambda^2-{1\over 4}\mu^2
-{1\over 2}v^2+v\mu)}.
\eqn\kergaussb$$
The redefinition
$$\tilde{K}(\lambda,\mu)\equiv e^{-{N\over 4}\lambda^2} K(\lambda,\mu)
e^{+{N\over 4}\mu^2},\eqn\redef$$
gives
$$\tilde{K}(\lambda,\mu)=\int_{-i\infty}^{+i\infty} {da\over 2\pi i}
\oint {dv\over 2\pi i} \prod_{l=0}^{N-1}\left({a-a_l \over v-a_l}\right)
{1\over a-v}
e^{N({1\over 2} a^2-a\lambda
-{1\over 2}v^2+v\mu)}.
\eqn\bhkerb$$
which coincides with eq.~$\bhker$ if one sets $a=it/N$. Notice that transformation
$\redef$ does not affect the value of determinants of the type $\detform$.

\chapter{Conclusion.}
We have derived determinant formulae for the correlation functions of eigenvalues
in terms of a kernel $K$ which has very simple algebraic properties:
it is a projector on a space of dimension $N$; its eigenvectors are known, even though
they are not as simple as in the $U(N)$-invariant case, since $K$ is no 
longer symmetric.

For the simple gaussian ensemble, the asymptotic 
behavior of the kernel $K(\lambda,\mu)$ when $\lambda-\mu$ is of order
$1/N$, $N\rightarrow\infty$,
has been known for a long time [\ME,\DY]:
$$K(\lambda,\mu)\sim {\sin s\over s},\quad s\sim N(\lambda-\mu).\eqn\asy$$
It is remarkable that in the general $U(N)$-invariant case [\BZ] as well
as in the gaussian case with an external source [\BH], this behavior remains
identical; combined with the determinant formulae, it
implies universal properties for
the correlation functions, as well as for the level spacing
distribution.
The latter quantity, first introduced by Wigner [\WI],
is empirically known to be universal for a wide range of models.
Here, it can be expressed in terms of $K$ using
the determinant formulae, and the relation takes a particularly simple
form in the domain of universality, that is
when one considers intervals of order $1/N$: if $p(s)$ is the level
spacing distribution, the spacing variable $\theta$ being defined by
$s=N\theta$, one has
$$p(s)={d^2\over ds^2} \det(1-\hat K)\eqn\lev$$
where $\hat K$ is the asymptotic form of $K$ in the
interval $[-\theta/2,\theta/2]$ as $N\rightarrow\infty$,
$\theta\rightarrow 0$, $s$ fixed (cf eq.~$\asy$), and $\det$
is the usual Fredholm determinant.
Thus, the universality of $K$ implies the short distance 
universality of the level spacing.
As opposed to the universal behavior
of the correlation function $\rho_2(\lambda,\mu)$
when $\lambda-\mu$ is of order 1,
which can be obtained by various standard large $N$ 
techniques [\AMB,\BZ,\BEEN], the short distance universal behavior
is more difficult to find,
and its generalization to an arbitrary potential with an external source
(using the newly found kernel $K$) is currently under study.

\medskip
\centerline{\bf Acknowledgement.}
I would like to thank E. Brezin, for introducing me to this problem and
for stimulating discussions.

\refout
\end